\UseRawInputEncoding
\documentclass[%
 reprint,
superscriptaddress,
 amsmath,amssymb,
 aps,
prb,
]{revtex4-1}

\usepackage{graphicx}
\usepackage{dcolumn}
\usepackage{bm}
\usepackage{amsmath,amssymb}
\usepackage{graphicx}
\usepackage{color}

\usepackage{hyperref}

\begin{document}

\title{A New Look at Calcium Digermanide CaGe$_2$: A High-Performing Semimetal Transparent Conducting Material for Ge Optoelectronics}

\author{V.M. Il'yaschenko}
\affiliation{Institute of Automation and Control Processes, Far Eastern Branch, Russian Academy of Science, 5 Radio Str., Vladivostok 690041, Russia}

\author{D.V. Pavlov}
\affiliation{Institute of Automation and Control Processes, Far Eastern Branch, Russian Academy of Science, 5 Radio Str., Vladivostok 690041, Russia}

\author{S.A. Balagan}
\affiliation{Institute of Automation and Control Processes, Far Eastern Branch, Russian Academy of Science, 5 Radio Str., Vladivostok 690041, Russia}

\author{A.V. Gerasimenko}
\affiliation{Institute of Chemistry FEB RAS, Vladivostok, 690022, Russia}

\author{N.V. Tarasenko}
\affiliation{B.I. Stepanov Institute of Physics, National Academy of Sciences of Belarus, Minsk, 220072, Belarus}

\author{Aleksandr A. Kuchmizhak}
\affiliation{Institute of Automation and Control Processes, Far Eastern Branch, Russian Academy of Science, 5 Radio Str., Vladivostok 690041, Russia}
\affiliation{Pacific Quantum Center, Far Eastern Federal University, Vladivostok, Russia}
\affiliation{Institute of Chemistry, Saint Petersburg State University, Saint Petersburg, 198504, Russia}

\author{Aleksandr V. Shevlyagin}
\email{shevlyagin@iacp.dvo.ru}
\affiliation{Institute of Automation and Control Processes, Far Eastern Branch, Russian Academy of Science, 5 Radio Str., Vladivostok 690041, Russia}
\affiliation{Institute of Chemistry, Saint Petersburg State University, Saint Petersburg, 198504, Russia}

\begin{abstract}
 Following a recently manifested guide of how to team up infrared transparency and high electrical conductivity within semimetal materials [C. Cui $et$ $al.$ Prog. Mater. Sci. 2023, 136, 101112], we evaluate an applicability of the calcium digermanide (CaGe$_2$) thin film electrodes for the advanced Ge-based optical devices. Rigorous growth experiments were conducted to define the optimal annealing treatment and thickness of the Ca-Ge mixture for producing stable CaGe$_2$ layers with high figure of merit (FOM) as transparent conducting material. Ab-initio electronic band structure calculations and optical modeling confirmed CaGe$_2$ semimetal nature, which is responsible for a demonstrated high FOM. To test CaGe$_2$ electrodes under actual conditions, a planar Ge photodetector (PD) with metal-semiconductor-metal structure was fabricated, where CaGe$_2$/Ge interface acts as Schottky barrier. The resulting Ge PD with semimetal electrodes outperformed commercially available Ge devices in terms of both photoresponse magnitude and operated spectral range. Moreover, by using femtosecond-laser projection lithography, a mesh CaGe$_2$ electrode with the relative broadband transmittance of 90\% and sheet resistance of 20 $\Omega$/sq. was demonstrated, which further enhanced Ge PD photoresponse. Thus, obtained results suggest that CaGe$_2$ thin films have a great potential in numerous applications promoting the era of advanced Ge optoelectronics.
\end{abstract}

\maketitle

\section{Introduction}
The modern civil, scientific and military applications raise new grand challenges for photonic and optoelectronic devices beyond scalability, low energy consumption and high photoelectric conversion yield, which are axiomatic golden standards. Among them flexible sensors, multicolor photodetectors (PDs) and their counterpart niche of the selectively blinds ones are worthy of special mention \cite{xie2017flexible,cao2020multicolor,xie2019recent}. These application fields are responsible not only for a rapid discovery of the novel light-sensitive materials at the cutting edge of the two-dimensional and topological condensed matter \cite{wu2021nanohybrid,yan2018toward,ezhilmaran2021recent,huo2022integrated,yao2017all,zhang2022weyl,liu2020semimetals,wang2017ultrafast}, but for an associated breakthrough of the transparent conducting materials (TCMs) \cite{moreira2022review,singh2022advanced,stoner2019chemical,liu2021mxenes}, which are essential parts of any optoelectronic device.

Transparent conducting oxides (TCOs) no longer could meet all mentioned requirements, despite a comprehensive list of the available materials \cite{spencer2022review, jaffray2022transparent} to replace the most utilized indium-tin-oxide with its high cost and brittleness \cite{he2016nanostructured}. In addition, widely used band engineering and doping close to the solubility limit \cite{cai2021perspective} hardly can result in simultaneous low optical losses and high electrical conductivity due to free-carriers absorption and impurity-enhanced electron scattering, respectively \cite{zhang2016correlated}. Owing to thickness reduction down to tens of nanometers, further accompanied by nanostructuring of all sorts, conventional thin metal films were granted a second chance \cite{zhang2022transparent, paeng2015low, zhang2021mask, qiao2023line, nam2019highly}. Unfortunately, there is a significant conductivity drop partially associated with island-like growth, which assumes needs for additional seeding layers deposition to improve electrical properties \cite{park2022resistivity,martinez2021ultrathin}, nothing to say about electrical losses induced by different metal film processing toward enhanced optical transparency resulted from defects introducing \cite{zhao2015stable}.
Quite different approach was introduced by Zhang $et.$ $al.$ (2016) with original materials screening based on high carrier effective mass rather than high concentration \cite{zhang2016correlated}. This concept allows plasma energy to be shifted deeply below visible range with much lower free-carriers absorption compared to conventional metals. The systems of interests became correlated metals and semimetals \cite{ok2021correlated, boileau2022highly}, which infrared transparency has been recently validated for the thin films of calcium disilicide (CaSi$_2$) and tungsten ditelluride (WTe$_2$) both possessing layered crystal structure and classified as trivial and type-II Weyl semimetals, respectively \cite{shevlyagin2022semimetal, cui2023strategies}. High intrinsic transparency in the optical telecommunication spectral range of the former material was shown in addition to a very low sheet resistance, which resulted in a record near-infrared (NIR) figure of merit (FOM) competitive with state-of-the-art TCOs and other TCMs. Moreover, a prototype photovoltaic device utilizing CaSi$_2$ top electrode instead of conventional metal-finger contacts was demonstrated resulting in the enhanced photovoltaic performance and clearly confirmed reliability of the semimetal approach \cite{shevlyagin2022textured}.

It is a common knowledge that Ge is a more attractive NIR optoelectronics platform in replacing Si owing to its almost twice-narrower band gap \cite{miller2009device}. That is why, developing the TCMs for that purpose is of high interest. Unfortunately, there are no reports on Ge integration with semimetals except for PDs with transparent electrodes made of graphene, \cite{kwon2022performance,jiang2022high} which is semimetal to a certain extent. However, as it is often the case for Me-IV alloys and compounds (Me = Ca, Mg; IV = Si, Ge, Sn), the solution lies on the surface. Thus, we propose to use calcium digermanide (CaGe$_2$) films in a similar way to CaSi$_2$ and Si. Despite the same crystallography (Zintl phases) \cite{beekman2019zintl}, CaGe$_2$ is less investigated in comparison with CaSi$_2$. Little is known about its electrical properties \cite{evers1974electrical} with no available data on optical investigations. Currently, CaGe$_2$ is used to produce 2D derivatives of Ge (germanene, germanane, polygermine etc.) by topochemical reaction \cite{jiang2016improved, vogg2000polygermyne,rosli2020siloxene, liu2019recent, chia2021functionalized} just like in the case of CaSi$_2$ and silicene, which is Si-based graphene analogue \cite{wang2022hundred,ryan2019silicene}.

In this work, we report on successful growth of the CaGe$_2$ films on the transparent insulating (Al$_2$O$_3$) and semiconducting (Ge) substrates. Ambient stability of the produced films depending on growth conditions was assessed by means of Raman spectroscopy and X-ray diffraction (XRD) methods. Based on electrical and optical measurements accompanied by first principles electronic band structure calculations and optical modeling, CaGe$_2$ was comprehensively characterized as a candidate for TCMs. It turned out, that among CaGe$_2$ polymorph modifications \cite{yaokawa2021polymorphic,yaokawa2018crystal}, only h2-CaGe$_2$ is stable under ambient conditions, while hR6-CaGe$_2$ tends to be formed as a primary phase in thin and/or annealed at low temperature Ca-Ge films. Next, it was shown that h2-CaGe$_2$ possesses semimetal behavior, which determines its high IR transparency and high conductivity similar to CaSi$_2$. After growth conditions optimization, CaGe$_2$ film demonstrates maximal optical transmittance close to 80\% at 2.25 µm wavelength, while its sheet resistance is as low as 13 $\Omega$/sq. In view of these features, CaGe$_2$ film in the form of finger electrodes was grown on Ge(001) substrate to produce Ge metal-semiconductor-metal (MSM) PD with back-to-back two CaGe$_2$/Ge Schottky barriers. The resulting Ge MSM PD demonstrates expanded photoresponse spectrum down to 2300 nm and enhanced sensitivity of 0.8 A/W under a small reverse bias compared with conventional Ge PD. Finally, to compensate for low transparency at optical frequencies, femtosecond (fs) laser perforation was used to produce a CaGe$_2$ mesh electrode. Even the highest perforation ratio of 92\% has only moderate influence on the resultant sheet resistance, while optical transparency in the (400-1000) nm range was enhanced by a factor of 3. As a result, CaGe$_2$ electrode could reach FOM of 0.75 $\Omega^{-1}$ that much higher or comparable with the currently applied near-IR and middle-IR TCMs \cite{gao2021p,khamh2018good,tong2016thermally,wang2019transparent,fukumoto2022ligand} and Ge PD with perforated CaGe$_2$ top electrode surpassed peak value photoresponse of 1 A/W at 1600 nm wavelength while overall improvement in photosensitivity for a wide photon band (400-2200) nm was confirmed. Of great importance is that either the semimetal nature of CaGe$_2$ or laser perforation have no crucial influence on Ge PD resulting response speed (tens of microseconds), which is comparable with other Ge-based heterojunction devices.

\section{Results and Discussion}
\subsection{Influence of growth conditions on ambient stability of the Ca-Ge layers}

 \begin{figure*}
\center{\includegraphics[width=0.65\linewidth]{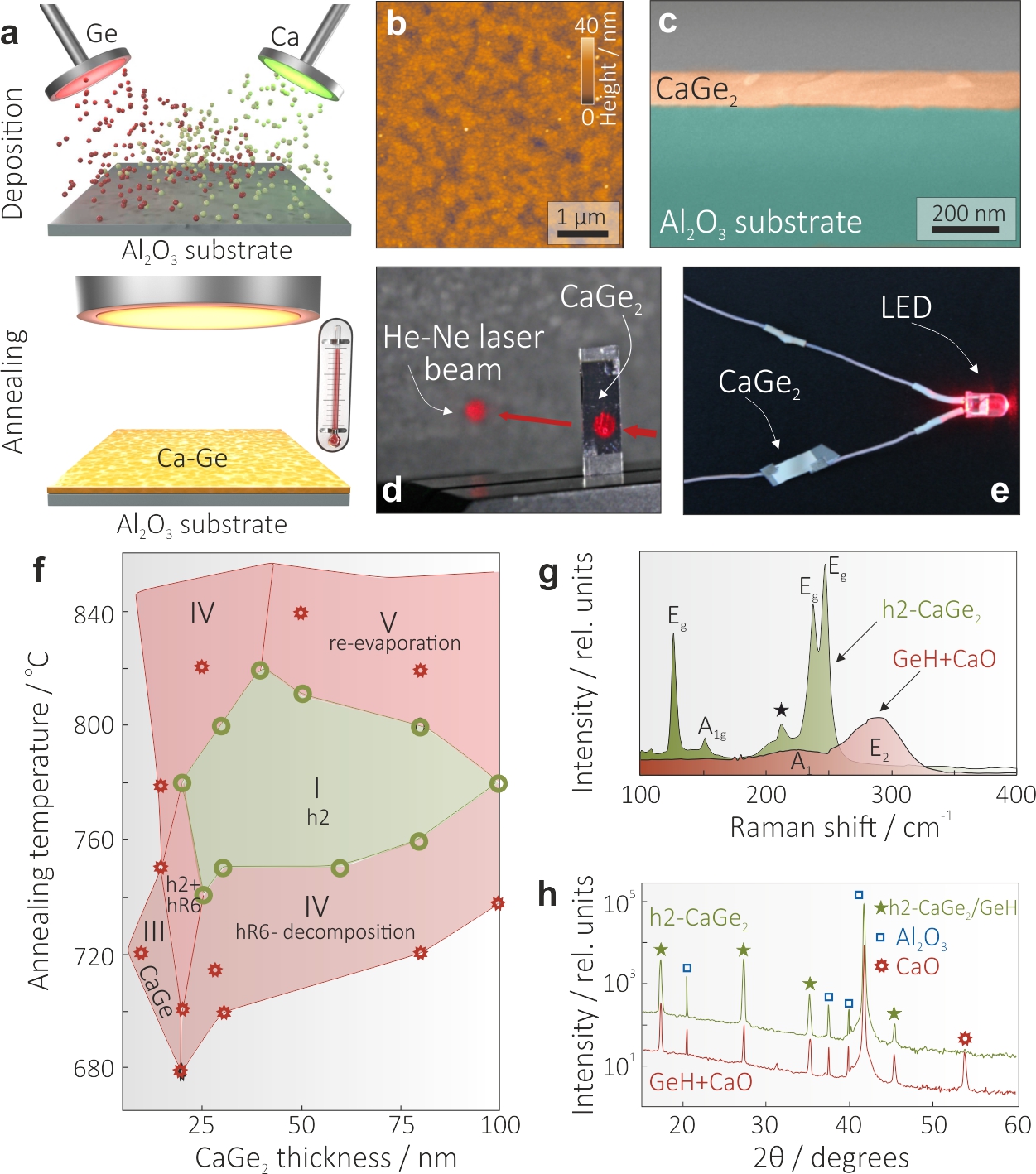}}
\caption{\space \textbf{Fabrication and characterization of the Ca-Ge layers on the Al$_2$O$_3$ substrate.} (a) Schematic sketch of the CaGe$_2$ growth procedure atop a sapphire wafer in a vacuum chamber. (b,c) Surface topography and cross-sectional morphology of 100-nm thick CaGe$_2$ film probed with AFM and SEM, respectively. (d,e) Photographs of the as-prepared CaGe$_2$ thin film on the Al$_2$O$_3$ arguing for its partial optical transparency at 633 nm and electrical conductance, respectively. (f) Conceptual Ca-Ge phase diagram derived from the growth conditions: annealing temperature and film thickness. (g,h) Representative Raman spectra and XRD patterns of the stable (green curves) and decomposed (pink curves) CaGe$_2$ thin films, respectively.}
\label{fig1}
\end{figure*}

We started our examinations from the attempts to grow CaGe$_2$ films on Al$_2$O$_3$ (sapphire) substrate. In doing so, a solid phase epitaxy (SPE) method was chosen suggesting some similarities between CaGe$_2$ thin films and CaSi$_2$ transparent conducting layers obtained previously \cite{shevlyagin2022semimetal}. Schematic illustration of the CaGe$_2$ SPE, which is a two-stage process, is pictured in Figure 1a. A set of the samples was obtained with varied Ca-Ge thickness (10-120) nm and annealing temperature (600-850)$^o$C. All grown films can be divided into two groups based on their stability, which is directly associated with phase composition driven by the growth conditions. Stable after air exposure and storage for 1 year samples demonstrate simultaneous partial optical transparency at least in the red spectral region (Figure 1d), sufficient electrical conductivity to supply the LED by passing a DC current through it (Figure 1e) and relatively smooth surface with RMS roughness not exceeding 3 nm in accordance with atomic force microscopy (AFM) data (Figure 1b). The latter is of great importance for developing TCMs with high FOM values, since pronounced surface relief could deteriorate both transparency and electrical conductivity, to say nothing of difficulties with materials processing towards real electrode engineering (lithography, patterning etc.). Scanning electron microscopy (SEM) investigations show that stable Ca-Ge layers consist of large grains and form continuous films (Figure 1c), while energy dispersive X-ray spectroscopy (EDX) confirms stoichiometry of CaGe$_2$ (Figure S1 in Supporting information). For unstable CaGe$_2$ films, in most cases their degradation took place within a few hours depending on its thickness except for the thinnest ones exhibiting some changes in their appearance obvious by a naked eye just after growth ending. The readers are further addressed to Supporting information for optical microscopy data and planar SEM images (Figure S2), which are helpful to express and preliminary examine the CaGe$_2$ films stability, while rigorous phase identification for the grown Ca-Ge layers probed with XRD and Raman method will be given below.

It is known that CaGe$_2$ can exist in several polymorph modifications, with hexagonal (space group \#186), two trigonal (space groups \#166 and \#164) and monoclinic (space group \#12) crystal lattices being the most frequently observed \cite{yaokawa2018crystal, yaokawa2021polymorphic, tobash2007synthesis}. The former two phases referred to as hR6 (or 6R) and h2 (or 2H) were experimentally observed including thin epitaxial films, while the latter two are metastable. In addition, it was reported that fluorine diffusion and crystal lattice stress release can promote stabilization of the other hexagonal (h4 or 4H) and trigonal (hR3 or 3R) CaGe$_2$ polymorphs. In this view, the Ca-Ge phase diagram is quite similar to the Ca-Si one in terms of the variety of phases and compounds. However, while both h3R and h6R trigonal CaSi$_2$ polymorphs are (i) stable, (ii) semimetal in nature and (iii) tolerant to degradation under ambient conditions, it is not the case for CaGe$_2$. Additional challenges arise, since h2- and hR6-CaGe$_2$ can be hardly separated by means of Raman spectroscopy and XRD examination \cite{vogg2000epitaxial,hara2021close,arguilla2017optical}. For example, Figure 1h represents a typical XRD pattern of the decomposed film obtained. The observed peaks can be attributed as diffraction from CaGe$_2$ and CaO crystal planes. However, it was further specified that GeH (germanane) and CaO formation took place rather than both hR6-CaGe$_2$ decomposition and h2-CaGe$_2$ preservation. The trick is that GeH and h2-CaGe$_2$ diffraction peaks are hardly distinguishable \cite{itoh2022crystal,pinchuk2014epitaxial}, which can be clearly seen by comparing the two XRD patterns corresponding to stable and decomposed Ge-Ca layers (Figure 1h). The only difference is in the observed CaO related peak in the latter case. The presence of the GeH in the decomposed film instead of the h2 phase is supported by Raman measurements presented in Figure 1g. The broad phonon bands centered at 275 cm$^{-1}$ (Ge-Ge bonds) and 225 cm$^{-1}$ (Ge-H bonds) can be categorically assigned to GeH (the pink graph) \cite{bianco2013stability}. This spectrum is in a marked contrast with that of measured from the stable CaGe$_2$ film (the green graph). At least four relatively narrow peaks were resolved, assigned as E$_g$ and A$_{1g}$ phonon modes of h2-CaGe$_2$. Thus, XRD measurements allowed detecting CaO presence, while Raman spectroscopy did the same for GeH. As a result, hR6-CaGe$_2$+H$_2$O$\rightarrow$CaO+2GeH can be tentatively suggested as an origin of the observed decomposition of thin CaGe$_2$ layers after air exposure. Concerning growth conditions, which lead to CaGe$_2$ decomposition, it can be stated that (i) thin ($<$15 nm) films show this tendency regardless annealing temperature, (ii) increase in total Ge-Ca thickness results in the presence of both h2 and hR6 phases and (iii) further increase in thickness opens a stability window for CaGe$_2$ in h2 modification with no hR6 traces, which suggest single-phase film growth. Too low annealing temperature results in CaGe formation (insulating phase), while too high temperature treatment does in re-evaporation of the deposited layers.

To outline some intermediate results on experiments of growth stable CaGe$_2$ films, we illustrated it by a schematic phase diagram shown in Figure 1f, which reflects chemical composition of the obtained samples addressed by Raman and XRD measurements in relation to growth conditions. In brief, it appears to be that a stable and single-phase h2-CaGe$_2$ film could not be grown thinner than 20 nm and below annealing temperature of 750$^o$C (region I). Outside the optimal ranges, the following processes make sense: phase coexistence (II), monogermanide formation (region III), total decomposition (region IV) and film re-evaporation (region V).

\subsection{Semimetal nature of the CaGe$_2$ and its influence on the obtained TCM figure of merit}

 \begin{figure}
\center{\includegraphics[width=0.95\linewidth]{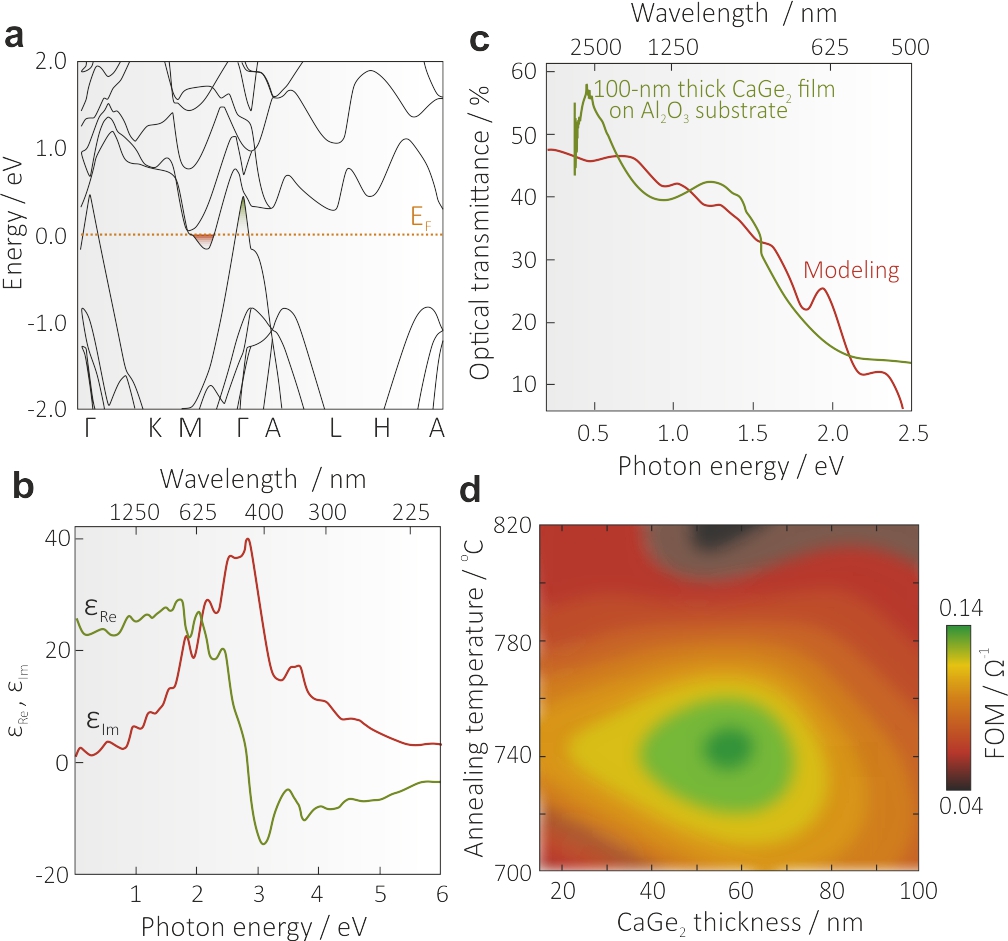}}
\caption{\space \textbf{Handling a puzzle of high IR transparency and high electrical conductance of CaGe$_2$ with theoretical modeling and experimental verification.} (a) \textit{Ab initio} electronic band structure of the bulk h2-CaGe$_2$ polymorph calculated with GGA. (b) Results of the Kramers-Kronig transformation for complex dielectric function of h2-CaGe$_2$. (c) Comparison of the measured and simulated optical transmittance of the 100 nm thick h2-CaGe$_2$ film grown on Al$_2$O$_3$ substrate. (d) FOM mapping of the Ca-Ge TCM layers versus its growth conditions.}
\label{fig2}
\end{figure}

After resolving the stability issue, optical and electrical properties of the h2-CaGe$_2$ films can be discussed from the view of both modeling and some practical aspects toward the TCMs performance. This screening was performed by first principles calculations, while extracted information is of high value in supporting optical and electrical measurements.
We started for the band structure calculations of the fully relaxed h2-CaGe$_2$ primitive cell and the resulting electron band diagram is shown in Figure 2a. There are multiple crossings at the Fermi level for both valence and electron bands. One can see at least one electron and one hole pockets above Fermi level (holes energy is counted inversely to that of for electrons) tinged with red and green, respectively. In addition, some peculiarities of the h2-CaGe$_2$ band structure are of great importance. First, CaGe$_2$ in h2 modification is a trivial indirect-type semimetal \cite{markov2019thermoelectric}, which means that electron and valence band extreme points located at the different k-points ($\Gamma$ and M, respectively) with the so-called negative band gap of about 0.6 eV (the sum of the electrons and holes pockets maximal energy offset with respect to Fermi level). Secondly, one can observe that with the exception of mentioned carriers pockets, there is a wide energy gap diving below and above Fermi level to about 0.6 and 0.3 eV, respectively. It resulted in a very low electron density of states (DOS) near Fermi levels compared to pure metals, but significantly larger than for semiconductors even under degeneracy doping. On the one hand, this electronic bands configuration allows demonstrating very low resistivity in the order of 10$^{-5}$-10$^{-4}$ $\Omega\cdot$cm for h2-CaGe$_2$ thin films (see Supporting information S3 for Hall measurements performed on the films of different thickness) compared with bulk single-crystals \cite{evers1974electrical}. From the other hand, low electronic DOS should results in the low optical DOS as well. Calculated complex dielectric function of the h2-CaGe$_2$ phase is plotted in Figure 2b and explains well the low optical losses in the IR spectral range up to 0.8 eV.

To provide an experimental verification of the ab initio calculations, optical transparency and sheet resistance were measured for 100-nm thick CaGe$_2$ film on Al$_2$O$_3$ substrate obtained under the growth conditions, which correspond to formation of a stable Ca-Ge phase (region I; Figure 1a). A typical experimental transmittance spectrum of the CaGe$_2$/Al$_2$O$_3$ sample is presented in Figure 2c together with the optical modeling results based on calculated complex dielectric function of the h2-CaGe$_2$ and the data available for Al$_2$O$_3$. In doing this, obtained dielectric constants within Fresnel law were used to calculate optical response of the CaGe$_2$/Al$_2$O$_3$ system assuming a smooth and sharp interface between materials and zero scattering. One can see a close agreement between theoretical and experimental evaluations. In addition, relatively thick (100 nm) CaGe$_2$ is semitransparent from the near-infrared (NIR) to middle-infrared (MIR) spectral ranges, while optical transparency of the CaGe$_2$/Al$_2$O$_3$ system reaches 60\% at 2500 nm.

 \begin{figure*}
\center{\includegraphics[width=0.85\linewidth]{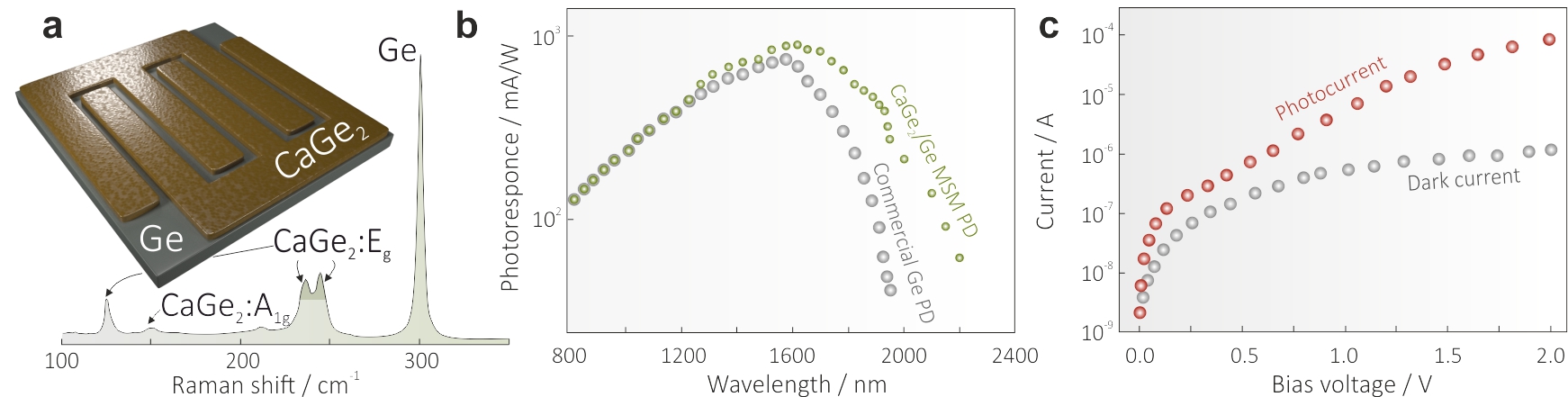}}
\caption{\space \textbf{Schottky-type Ge PD with transparent CaGe$_2$ electrodes (CaGe$_2$/Ge PD).} (a) Typical Raman spectra of the CaGe$_2$ deposited onto Ge substrate confirming technology compatibility and stability of the grown layer. The inset shows a schematic view of the planar Schottky-type CaGe$_2$/Ge PD. (b) Current-voltage characteristics of the CaGe$_2$/Ge PD under the dark and illuminated conditions (@1550 nm). (c) Room-temperature photoresponse spectra of the CaGe$_2$/Ge PD and commercial one. Both devices were driven at the reverse bias voltage of 2V.  }
\label{fig3}
\end{figure*}

Next, rigorous investigations were performed to define CaGe$_2$ growth conditions (thickness and annealing temperature), which lead to the highest FOM value in accordance with the following expression: FOM = -1/[R$_{sheet}\cdot$ln(T)], where R$_{sheet}$ and T are sheet resistance and optical transmittance (at the selected wavelength or averaged over the specific range), respectively \cite{gordon2000criteria}. The results of this screening is shown as 3D mapping in Figure 2d. It was found that the transparency window of the h2-CaGe$_2$ itself is located in the near-IR (1-3 $\mu$m) spectral range. The measured sheet resistance varied from 8 to 30 $\Omega$/sq. depending on both film thickness and annealing conditions. After all, 60 nm thick CaGe$_2$ film annealed at 750$^o$C demonstrated the highest relative optical transparency of 78\% at 2.25 $\mu$m and moderate sheet resistance of 16 $\Omega$/sq. These parameters give the FOM values of the 0.33 and 0.13 $\Omega^{-1}$ at the wavelength of maximal transmittance (2250 nm) and for an transparency averaged over (400-7000) nm range, respectively. Obtained FOM value for the h2-CaGe$_2$ is competitive with other state-of-the-art TCMs currently applied for NIR-MIR applications \cite{gao2021p,khamh2018good,tong2016thermally,wang2019transparent,fukumoto2022ligand}.

\subsection{CaGe$_2$/Ge planar MSM NIR-SWIR photodetector: fabrication and characterization}

To test h2-CaGe$_2$ TCM under the real photovoltaic device operations, we grew a corresponding thin film on the Ge(111) substrate instead of Al$_2$O$_3$ transparent one using the same optimized conditions found previously (60 nm thick CaGe$_2$ layer annealed at 750$^o$C). The deposition and annealing of the Ge-Ca bilayer was done through a tantalum mask for in situ formation of the two CaGe$_2$ pads. Thus, we obtained a MSM photodetector acting as two back-to-back Schottky junctions with a schematic design pictured in Figure 3a. Raman measurements suggest a good crystallinity of the grown semimetal layers with observed phonon bands corresponding only to CaGe$_2$ and Ge with no traces of film decomposition and other Ca-Ge alloys. Figure 3b demonstrates dark I-V characteristic of the planar MSM structure with clear Schottky behavior. Under the assumption of the thermionic-field emission regime \cite{katz2001gain}, a Schottky barrier height was calculated to be 0.5 eV, while reverse saturation current is equal to 50 nA. Under laser illumination at 1550 nm, CaGe$_2$/Ge/CaGe$_2$ MSM diode demonstrates no photocurrent generation under zero-bias condition. However, even a small bias voltage results in pronounced photocurrent, wherein current flow is enhanced by two orders of magnitude at the applied 2V under illumination compared to dark conditions. Thus, a typical operation of the MSM PD with symmetrical Schottky contacts was confirmed. Next, a room temperature photoresponse at the bias voltage of 2V was measured for the CaGe$_2$/Ge/CaGe$_2$ MSM PD. Results obtained are plotted in Figure 3c together with characteristics of the commercial Ge PD. It is obvious that application of the CaGe$_2$ transparent electrodes resulted in the spectral range expanded into the shortwave IR region below the band gap of Ge (0.67 eV). We estimated a photoelectric threshold energy to be 0.54 eV (2300 nm), which well correlates with Schottky barrier height obtained from the I-V curve. Moreover, there is the photoresponse enhancement in the (1200-1900) nm range owing to a high transparency of the CaGe$_2$ electrodes, which result in higher photogeneration occurring in Ge. Thus, a simple but effective approach of advancing Ge-based photodiode performance in terms of maximal sensitivity and red shifting of the cut-off wavelength is demonstrated.

\subsection{Improvement of the CaGe$_2$ TCM performance with laser projection lithography}

 \begin{figure*}
\center{\includegraphics[width=0.85\linewidth]{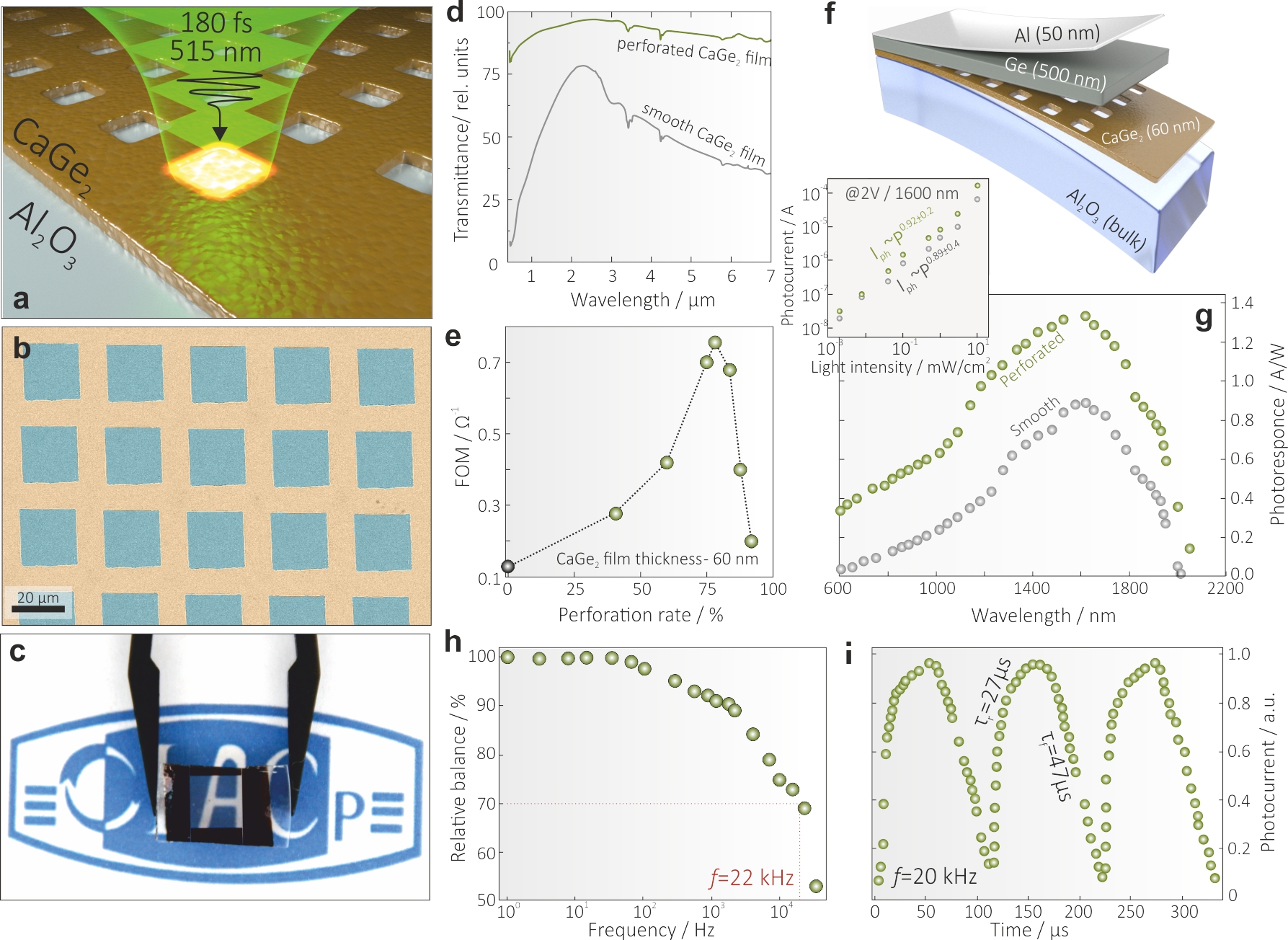}}
\caption{\space \textbf{Advanced Ge PDs with laser-perforated CaGe$_2$ electrodes.} (a) Schematic illustration of fs-laser projection lithography for micropatterning of the CaGe$_2$ film on the Al$_2$O$_3$ substrate. (b) SEM image of 60-nm thick CaGe$_2$ film patterned with square-shaped openings forming a mesh electrode. (c) Photograph of the CaGe$_2$ film with a 4$\times$4 mm$^2$ laser-patterned area confirming its high visible transparency. (d) Relative transmittance of the CaGe$_2$/Al$_2$O$_3$ sample before (smooth film) and after laser patterning (film with perforation ratio of 78\%). Both spectra were normalized over the transmittance of the Al$_2$O$_3$ substrate. (e) FOM of the 60-nm thick CaGe$_2$ film as a function of the perforation rate. (f) A schematic view of the vertical Schottky-type Ge PD with perforated CaGe$_2$ electrode. (g) Photoresponse performance of the Schottky-type Ge PDs with smooth and perforated CaGe$_2$ electrodes. Both curves were obtained at room temperature and reverse bias voltage of -2V. The inset demonstrates linear dependence of the photocurrent on the incident light intensity (@1550 nm) for both Ge PDs. (h) The relative balance 100\%$\times$(I$_{ph}^{max}$ – I$_{ph}^{min})$/I$_{ph}^{max}$ versus switching frequency, showing a 3 dB cutoff frequency of 22 kHz for Ge PD with a perforated CaGe$_2$ electrode. (i) Normalized operation cycles of the Ge PD measured under NIR illumination (@1550 nm) modulated  at 20 kHz for determining both rise ($\tau_{rise}$) and fall times ($\tau_{fall}$). }
\label{fig4}
\end{figure*}

Despite a high NIR-MIR TCM performance, the grown CaGe$_2$ films have low optical transparency in the visible spectral range. Fortunately, it can be tuned by making a perforated mesh electrode by direct fs-laser patterning. For this purpose, 60 nm thick CaGe$_2$ film was processed using a flat-top square-shape laser beam as schematically illustrated in Figure 4a. The laser fluence was kept slightly above the single-pulse ablation threshold of the corresponding CaGe$_2$ film (see detailed description of the ablation threshold measurements in Supporting information - Figure S4). Typical morphology of the 60-nm thick CaGe$_2$ film grown on Al$_2$O$_3$ substrate after laser processing is demonstrated by top-view SEM image in Figure 4b, with the photograph of the resulted Al$_2$O$_3$/CaGe$_2$ sample after laser perforation clearly demonstrating the enhanced optical transparency compared to non-perforated areas (Figure 4c). The sidewalls of the CaGe$_2$ mesh are smooth and no traces of the ejected submicron-sized particles can be found, which suggest high suitability of the CaGe$_2$ material for laser patterning. It is important to note, that a maximal rim height of the square-shape openings was about 90 nm (Figure S5), which is only 1.7 higher than initial CaGe$_2$ film thickness indicating that under chosen laser patterning conditions the outward radial mass transfer of the molten Ca-Ge is not much intense.

Representative optical transparency spectra of the 60-nm thick CaGe$_2$ film before and after laser perforation are presented in Figure 4d. The most pronounced changes are observed in the (400-1500) nm range, which are characterized by 3 times higher transparency of the perforated sample in comparison to untreated one, while transparency in the (1500-7000) nm range are intrinsically high for CaGe$_2$ material owing to its semimetal nature. Raman measurements confirmed (not shown) that even at high perforation ratio, laser ablation does not strongly affect crystallinity of the remaining sections of the CaGe$_2$ films. Compared with continuous films, a mesh CaGe$_2$ electrode with an optimal perforation ratio (determined to be 78\%) demonstrates six-fold increases in the FOM value reaching 0.75 $\Omega^{-1}$ (see Figure 4e) due to an optical transparency averaged over (400-7000) nm beating 90\% after laser processing. The lower perforation ratio is not enough to reach high optical transparency. On the contrary, extensive perforation rate above 80\% decreases FOM value owing to drastically increased sheet resistance up to 32 $\Omega$/sq., which is 3 times higher compared with initial smooth CaGe$_2$ film, however being still acceptable for TCM applications.

Finally, to confirm and highlight the advantage of fs-laser projection lithography in tuning optical properties of CaGe$_2$ film, a vertical Ge PD was fabricated on a sapphire substrate with a perforated transparent electrode (face contact) and thin Al film acting as a back electrode. The schematic view of the resulting vertical Ge Schottky PD is shown in Figure 4f assuming light irradiation from the CaGe$_2$ side. As the most illustrative test, we next directly compared the photoresponse characteristics of the two Ge PDs with a flat and perforated CaGe$_2$ electrodes (Figure 4g). One can see that the latter device generally follows the trend observed for the optical transmittance measurement. For instance, integrated over (600-2000) nm range photoresponse and peak photoresponse value (at 1600 nm) were enhanced by 85\% and 44\% (the room-temperature photoresponse exceeds 1 A/W under -2V bias voltage), respectively, while an expanded operation spectral range was observed for the Ge device with perforated top electrode. Of great importance, integrating the Ge PD platform with CaGe$_2$ electrodes has no negative influence on the light detection, which is linearly dependent from the irradiation intensity regardless of CaGe$_2$ electrode type (see inset in Figure 4g). In addition, we investigated the response speed of the device with a perforated electrode. The relative balance of the Ge PD device versus a switching frequency of pulsed IR light irradiation (@1550 nm laser) is presented in Figure 4h with a relatively high 3dB frequency of 22 kHz, which claims for Ge PD with transparent conducting electrode made from semimetal film to be capable of detecting fast optical signals. The response speed was further evaluated by analyzing the rising and falling edges of the photoresponse time curve under a cycle conditions, which are normally calculated as the time intervals for the response to rise from 10\% to 90\% and vice versa (Figure 4i). A fast rise/fall time ($\tau_r$/$\tau_f$) of 27/47 $\mu$s was obtained. We could attribute this relatively quick response speed to the low density of trap centers at the Ge/CaGe$_2$ heterointerface, which is additionally confirmed by the photoresponse versus incident light dependence, and a very effective separation of photogenerated carriers by the built-in electric field formed at the CaGe$_2$/Ge Schottky junction area. To sum up, the critical parameters for the developed vertical Ge PD with Schottky type contact made of perforated film of semimetalic CaGe$_2$ were compared with other Ge Schottky type PDs \cite{kwon2022performance,jiang2022high,falcone2022graphene,huang2021high,huang2018low,cho2016resonant,an2022black,dushaq2017metal,ang2008novel,zumuukhorol2019variations,yin2022self,chang2019high,zeng2013monolayer,kim2021highly,yang2017ultrathin,luo2019pdse2,zhao2020interface,xiong2022mxene} with an emphasis on MSM structures and listed in Table 1.

\begin{table*}[]
\begin{tabular}{ccccccc}
\hline
\multicolumn{1}{|c|}{\textbf{Structure}}                                                                             & \multicolumn{1}{c|}{\textbf{\begin{tabular}[c]{@{}c@{}}Driving \\ voltage\end{tabular}}} & \multicolumn{1}{c|}{\textbf{\begin{tabular}[c]{@{}c@{}}Wavelength\\ range\end{tabular}}} & \multicolumn{1}{c|}{\textbf{\begin{tabular}[c]{@{}c@{}}Maximal\\ photoresponse\end{tabular}}} & \multicolumn{1}{c|}{\textbf{\begin{tabular}[c]{@{}c@{}}Dark \\ current\end{tabular}}} & \multicolumn{1}{c|}{\textbf{\begin{tabular}[c]{@{}c@{}}Rise and fall\\ time\end{tabular}}} & \multicolumn{1}{c|}{\textbf{Reference}} \\ \hline
\textbf{\begin{tabular}[c]{@{}c@{}}Perforated \\ CaGe$_2$/Ge/Al\\ MSM PD\end{tabular}}                               & \textbf{2 V}                                                                             & \textbf{0.6 - 2 $\mu$m}                                                                  & \textbf{\begin{tabular}[c]{@{}c@{}}1.33 A/W \\ @1600 nm\end{tabular}}                         & \textbf{0.31 mA/cm$^2$}                                                               & \textbf{\begin{tabular}[c]{@{}c@{}}27 $\mu$s/ \\ 47 $\mu$s\end{tabular}}                   & \textbf{This work}                      \\
\begin{tabular}[c]{@{}c@{}}Ge-Si p-i-n \\ PD with \\ graphene \\ electrode\end{tabular}                              & 2 V                                                                                      & 1.3 - 1.8 $\mu$m                                                                         & \begin{tabular}[c]{@{}c@{}}0.13 A/W \\ @1550 nm\end{tabular}                                  & 2 mA/cm$^2$                                                                           & -                                                                                          & {[}63{]}                                \\
\begin{tabular}[c]{@{}c@{}}Ge/ITO\\ Schottky PD\end{tabular}                                                         & 1 V                                                                                      & \begin{tabular}[c]{@{}c@{}}selected \\ wavelength\end{tabular}                           & \begin{tabular}[c]{@{}c@{}}0.13 A/W \\ @1550 nm\end{tabular}                                  & 33 mA/cm$^2$                                                                          & -                                                                                          & {[}64{]}                                \\
\begin{tabular}[c]{@{}c@{}}Ge/ITO/Au \\ Schottky PD\end{tabular}                                                     & 0.2 V                                                                                    & 0.8-1.65 $\mu$m                                                                          & \begin{tabular}[c]{@{}c@{}}0.62 A/W \\ @1310 nm\end{tabular}                                  & 1.4 mA/cm$^2$                                                                         & -                                                                                          & {[}65{]}                                \\
\begin{tabular}[c]{@{}c@{}}Ge nano- \\ membrane MSM PD \\ with resonant cavity\end{tabular}                          & 1 V                                                                                      & selected wavelength                                                                      & \begin{tabular}[c]{@{}c@{}}0.18 A/W \\ @1550 nm\end{tabular}                                  & 100 mA/cm$^2$                                                                         & -                                                                                          & {[}66{]}                                \\
\begin{tabular}[c]{@{}c@{}}Ge Schottky \\ PD with \\ doped graphene\end{tabular}                                     & 1 V                                                                                      & selected wavelength                                                                      & \begin{tabular}[c]{@{}c@{}}1.27 A/W \\ @1550 nm\end{tabular}                                  & 63 mA/cm$^2$                                                                          & -                                                                                          & {[}35{]}                                \\
\begin{tabular}[c]{@{}c@{}}Reactive ion etched\\  Ge MSM \\ with Ti/Au electrodes\end{tabular}                       & 0.5 V                                                                                    & 1.5-2 $\mu$m                                                                             & \begin{tabular}[c]{@{}c@{}}2 A/W \\ @1550 nm\end{tabular}                                     & 0.616 mA                                                                              & -                                                                                          & {[}67{]}                                \\
\begin{tabular}[c]{@{}c@{}}Ge on Si \\ MSM PD\end{tabular}                                                           & 1 V                                                                                      & selected wavelength                                                                      & \begin{tabular}[c]{@{}c@{}}0.42 A/W \\ @1310 nm\end{tabular}                                  & 0.76 mA/cm$^2$                                                                        & -                                                                                          & {[}68{]}                                \\
\begin{tabular}[c]{@{}c@{}}NiGe/Ge \\ Schottky PD\end{tabular}                                                       & 1 V                                                                                      & selected wavelength                                                                      & \begin{tabular}[c]{@{}c@{}}0.36 A/W \\ @1550 nm\end{tabular}                                  & 100 mA/cm$^2$                                                                         & 15 GHz                                                                                     & {[}69{]}                                \\
\begin{tabular}[c]{@{}c@{}}Ge MSM PD \\ with Pt electrodes\end{tabular}                                              & 2 V                                                                                      & 1.53-1.61 $\mu$m                                                                         & \begin{tabular}[c]{@{}c@{}}0.41 A/W \\ @1550 nm\end{tabular}                                  & 2 mA/cm$^2$                                                                           & -                                                                                          & {[}70{]}                                \\
\begin{tabular}[c]{@{}c@{}}Ge Schottky PD \\ with graphene \\ and Au electrodes\end{tabular}                         & 1V                                                                                       & 1.064-1.85 $\mu$m                                                                        & \begin{tabular}[c]{@{}c@{}}1.82 A/W \\ @1064 nm\end{tabular}                                  & 1.6 mA/cm$^2$                                                                         & -                                                                                          & {[}36{]}                                \\
\begin{tabular}[c]{@{}c@{}}Ge heterojunction \\ PD with \\ topological insulator \\ Bi$_2$Te$_3$\end{tabular}        & zero bias                                                                                & selected wavelength                                                                      & \begin{tabular}[c]{@{}c@{}}0.97 A/W \\ @1064 nm\end{tabular}                                  & -                                                                                     & 12.1 $\mu$s                                                                                & {[}71{]}                                \\
\begin{tabular}[c]{@{}c@{}}Ge/ZnO heterojunction\\ PD with transparent \\ graphene electrode\end{tabular}            & 10 V                                                                                     & 0.3-1.8 $\mu$m                                                                           & \begin{tabular}[c]{@{}c@{}}0.75 A/W \\ @1550 nm\end{tabular}                                  & -                                                                                     & 250 Hz                                                                                     & {[}72{]}                                \\
\begin{tabular}[c]{@{}c@{}}Ge Schottky PD \\ with graphene   \\ monolayer\end{tabular}                               & zero bias                                                                                & 1.2-1.6 $\mu$m                                                                           & \begin{tabular}[c]{@{}c@{}}0.052 A/W \\ @1400 nm\end{tabular}                                 & -                                                                                     & 23 $\mu$s/108 $\mu$s                                                                       & {[}73{]}                                \\
\begin{tabular}[c]{@{}c@{}}Ge Schottky PD \\ with graphene   \\ electrode and \\ Al$_2$O$_3$ interlayer\end{tabular} & 2 V                                                                                      & 0.5-1.65 $\mu$m                                                                          & \begin{tabular}[c]{@{}c@{}}1.2 A/W \\ @1310 nm\end{tabular}                                   & -                                                                                     & -                                                                                          & {[}74{]}                                \\
\begin{tabular}[c]{@{}c@{}}Ge/graphene \\ Schottky PD\end{tabular}                                                   & 1 V                                                                                      & 0.35-1.65 $\mu$m                                                                         & \begin{tabular}[c]{@{}c@{}}66 A/W \\ @532 nm\end{tabular}                                     & -                                                                                     & 5.6 ms/3.5 ms                                                                              & {[}75{]}                                \\
\begin{tabular}[c]{@{}c@{}}Ge nanocones \\ covered with   \\ semimetal PdSe2\end{tabular}                            & zero bias                                                                                & selected wavelength                                                                      & \begin{tabular}[c]{@{}c@{}}0.53 A/W \\ @1550 nm\end{tabular}                                  & -                                                                                     & 25.4 $\mu$s/38.5 $\mu$s                                                                    & {[}76{]}                                \\
\begin{tabular}[c]{@{}c@{}}3D-\\ graphene/2D-\\ graphene/Ge\end{tabular}                                             & 1 V                                                                                      & selected wavelength                                                                      & \begin{tabular}[c]{@{}c@{}}1.7 A/W \\ @1550 nm\end{tabular}                                   & -                                                                                     & 68 $\mu$s/70 $\mu$s                                                                        & {[}77{]}                                \\
\begin{tabular}[c]{@{}c@{}}MXene/Ge \\ Schottky PD\end{tabular}                                                      & zero bias                                                                                & 0.35-1.55 $\mu$m                                                                         & \begin{tabular}[c]{@{}c@{}}3.14 A/W \\ @1550 nm\end{tabular}                                  & -                                                                                     & 1.4 $\mu$s/4.1 $\mu$s                                                                      & {[}78{]}
\end{tabular}
\end{table*}

As can be clearly seen, such phenomena as transparent conducting electrodes have rarely been addressed concerning the applications in Ge PDs except for some works focusing on graphene, ITO and MXenes integration, nothing to say about much fewer reports on Ge/semimetal or Ge/topological insulator heterojunction PDs. The current work demonstrates that low dark current, high responsivity in wide wavelength range and fast response speed is achievable by introducing semimetal CaGe$_2$ transparent conducting electrode into Ge optoelectronics. In particular, spectral operation range of the proposed Ge PD is the broadest than that of other photodetectors of the same type (Schottky or MSM). Concerning both photoresponsivity and operation speed, fabricated PD is obviously outperform commercially available Ge PD, while demonstrating competitive characteristics even compared to black Ge based PD. Moreover, a self-powered operation (zero bias driving voltage) is not anyhow restricted by CaGe$_2$ application and could be potentially achievable by the widely known "symmetry breaking" approach \cite{wang2020breaking,zhou2018self} for Schottky type PDs, which could be a question of future work.  Thus, obtained results suggest that laser-perforated CaGe$_2$ thin films have a great potential as transparent conducting materials for Ge optoelectronic devices and applications.

\section{Conclusion}

In the present work, a CaGe$_2$ commonly used in chemical reactions toward producing 2D Ge structures and derivatives was examined as a candidate for transparent conducting material to be applied in Ge-based optoelectronics. Obtained CaGe$_2$ films demonstrated electrical and optical properties quite similar to its vis-a-vis CaSi$_2$. Semimetal topology of the electronic bands provided high both optical transparency and electrical conductivity, which were suggested to be the common features of the Ca-IV$_2$ (IV=Si, Ge) Zintl phases. However, experimental investigations revealed some critical differences in stability and optical performance of the CaGe$_2$ layers. The former is resulted in relatively narrow ranges of the growth conditions, which prevent ambient induced decomposition and promote a single-phase CaGe$_2$ film formation. The latter is in the redshift of the transparency window to the MIR region compared to CaSi$_2$ film of equal thickness. Nevertheless, after finding the optimal growth conditions, CaGe$_2$ layers were used to produce planar metal-semiconductor-metal photodetector with two back-to-back Ge/CaGe$_2$ Schottky junctions. The resulting Ge MSM PD with CaGe$_2$ top transparent electrodes outperformed commercial Ge PD in both absolute photoresponse value operated spectral range. Next, a low intrinsic visible transparency of the CaGe$_2$ film was improved by a film perforation with fs-laser projection lithography. A mesh electrode with perforation ratio of ~80\% reaches FOM value of 0.75 $\Omega^{-1}$ in the wide spectral range of (400-7000) nm, which makes it competitive with other TCMs. Finally, patterned CaGe$_2$ thin film was used towards advanced Ge PDs developing, which demonstrated low dark current, high responsivity in wide wavelength range and fast response speed compared to state-of-the-art Ge-based optoelectronic devices.

Of great importance is that we demonstrated a new application of the CaGe$_2$ beyond just using it as precursor material for germanane, germanene and their polymorphs and derivatives production, while results obtained could become a great step toward mass Ge optoelectronics.

\section{Experimental Section}

All CaGe$_2$ thin films were grown in a turbo-pumped vacuum chamber with a base pressure of 10$^{-6}$ mbar. Solid phase epitaxy was used to form CaGe$_2$ layers in a way similar to CaSi$_2$ growth, which can be found elsewhere \cite{shevlyagin2022semimetal}. First, a Ge-Ca bilayer (K-cells, evaporation rates of 25 nm/min) was deposited onto substrate kept at room temperature followed by thermal annealing.

Phase composition and crystallinity of all samples were examined by Raman spectroscopy setup (NTEGRA SPECTRA II) and X-ray diffraction method (XRD) working in 2$\Theta$/$\omega$ mode with Cu K$_{\alpha}$ radiation source, parallel beam optics (Rigaku SmartLab). In addition, visual control of the sample surface modification after air exposure was carried out with an optical microscopy.

Combination of the four-point probe setup for sheet resistance measurements (Teslatron PV) and Fourier Transform Infrared spectrometer (Bruker Vertex V80) coupled to the IR microscope (Bruker Hyperion 2000) for optical properties evaluation were used to calculate FOM values of Ca-Ge films as TCM using the following expression: FOM = -1/[R$_{sheet}\cdot$ln(T)], where R$_{sheet}$ and T are sheet resistance and optical transmittance (at the selected wavelength or averaged over the specific range), respectively \cite{gordon2000criteria}.

For a deeper insight into semimetal electronic structure of the CaGe$_2$ and checking its optical properties, first principles calculations and multilayer optical modeling were performed. Further particulars of the calculations are as follows. The density-functional theory (DFT) \cite{hohenberg1964density,kohn1965self} calculations were performed with the package VASP \cite{kresse1996efficiency,kresse1996efficient}. The generalized-gradient approximation (GGA) to the exchange-correlation functional was used \cite{perdew1996generalized}.  Non-spherical contributions from the gradient corrections were included \cite{adler1962quantum,wiser1963dielectric}. The cut-off energy of 600 eV and gamma-centered k-points mesh of 20?20?8 were used. Break condition for self-consistency loop was of 10$^{-8}$ eV. Convergence with respect to cut-off energy and k-points density was performed. Calculation of frequency-dependent dielectric constants is implemented in VASP code as follows: the imaginary part of dielectric tensor $\epsilon_2^{\alpha\beta}(\omega)$ is determined by a summation over empty band states using the equation \cite{gajdovs2006linear}:

\begin{equation}
\epsilon_2^{\alpha\beta}(\omega) = \dfrac{2\pi e^2}{\Omega \epsilon_0} \sum\limits_{\kappa \nu c} \delta (E_\kappa ^c - E_\kappa ^{\nu} - \hbar\omega) \mid \langle \Psi_{\kappa}^c \mid  \textbf{u}\cdot\textbf{r} \mid \Psi_{\kappa}^{\nu} \rangle \mid ^2,
 \end{equation}

where $\epsilon_0$ is the vacuum dielectric constant, $\Omega$ is the volume, $\nu$ and $c$ represents the valence and conduction bands respectively, $\hbar\omega$ is the energy of the incident phonon, \textbf{u} is the vector defining the polarization of the incident electric field, \textbf{u}$\cdot$\textbf{r} is the momentum operator, and $\Psi_\kappa^c$ and $\Psi_\kappa^{\nu}$ are the wave functions of the conduction and valence band at the k point, respectively. The real part of dielectric tensor is obtained by the Kramers-Kronig relation:

\begin{equation}
\epsilon_1^{\alpha\beta}(\omega) = 1 + \dfrac{2}{\pi} P \int_0^\infty \dfrac{\epsilon_2^{\alpha\beta}(\omega^{'})\omega^{'}}{\omega^{'2} - \omega^{2} + i \eta} d\omega^{'},
 \end{equation}

where P denotes the principle value. Obtained dielectric constants and Fresnel law were used to calculate reflecting and refracting layers' response on radiation assuming a smooth and sharp interface between materials and zero scattering.

For the device applications, a CaGe$_2$ layer of optimal thickness and thermal annealing was grown on Ge(001) and Al$_2$O$_3$(0001) substrates to fabricate Ge planar and vertical Schottky MSM photodetectors. The latter CaGe$_2$ film was delicately perforated using a fs-laser projection lithography with second-harmonic (515 nm) 180 fs laser pulses generated by regeneratively amplified Yb:KGW system at 50 KHz maximal repetition rate. The output Gaussian laser beam was shaped to flat-top square-shape beam with the lateral size of 25x25 $\mu$m$^2$ to perform uniform laser ablation of square-shape surface area of the films. To obtain such intensity distribution in the focal plane of the dry microscope objective with a numerical aperture (NA) of 0.42, the output laser beam was expanded, while its central part with nearly uniform intensity profile was passed through the square-shape pinhole drilled in the aluminum foil by direct laser ablation \cite{zhizhchenko2020light}. A 4$f$ optical system consisting of a lens and a focusing objective was further used to project with magnification the uniform square-shape intensity profile at the output of the pinhole to the focal plane of the objective. Laser perforation was followed by p-Ge (500 nm) and Al (50 nm) films deposition to complete vertical photodetector structure.

The room-temperature current-voltage (I-V) and photoresponse spectra of the resulting planar MSM PD with CaGe$_2$/Ge Schottky barriers were measured in a standard manner described elsewhere \cite{shevlyagin2020probing}. To detect the response speed of the resulted Ge PD, pulsed optical signals with varied frequencies were produced by modulation (Thorlabs optical chopper) of monochromatic light with 1550 nm wavelength (Hamamatsu Xe lamp and monochromator Solar Tii, MS3504i), while the output photocurrent was recorded with an oscilloscope (DPO2012B, Tektronix).

\section{Acknowledgements}

The work was supported by the Russian Science Foundation under the Grant \#23-49-10044 (https://rscf.ru/project/23-49-10044/). Convergence verification in k-points density and cut-off energy for ab initio calculations were performed on HPC-cluster "Akademic V.M. Matrosov" (Irkutsk Supercomputer Center of SB RAS, https://hpc.icc.ru). Calculation of frequency-dependent dielectric constants was performed using the equipment of Shared Resource Center "Far Eastern Computing Resource" (Shared Resource Center "Far Eastern Computing Resource" IACP FEB RAS, https://cc.dvo.ru).

\section{Conflict of Interest}
The authors declare no conflict of interest.

\end{document}